\documentclass[final,5p,times]{elsarticle}

\usepackage{amssymb}

\usepackage{ifthen}
\usepackage{color}
\usepackage{proof}

\definecolor{rot}{rgb}{0,0,0}
\definecolor{dunkelgruen}{rgb}{.133,0.545,0.133}
\newcommand{\add}[1]{\textcolor{rot}{\textrm{#1}}}
\newif\ifcom
\newif\ifdel
\comtrue               %
\delfalse

\journal{Applied Surface Science}

\begin{document}

\begin{frontmatter}



\title{Magnetically Ordered Insulators for Advanced Spintronics}


\author[WMI,TUM]{Matthias Althammer}
\ead{matthias.althammer@wmi.badw.de}
\author[TUD,CTEM]{Sebastian T.~B.~Goennenwein}

\author[WMI,TUM,NIM]{Rudolf Gross}

\address[WMI]{Walther-Mei{\ss}ner-Institut, Bayerische Akademie der Wissenschaften, 85748 Garching, Germany}
\address[TUM]{Physik-Department, Technische Universit\"at M\"unchen, 85748 Garching, Germany}
\address[TUD]{Institut f\"ur Festk\"orperphysik, Technische Universit\"at Dresden, 01062 Dresden, Germany}
\address[CTEM]{Center for Transport and Devices of Emergent Materials, Technische Universit\"at Dresden, 01062 Dresden, Germany}
\address[NIM]{Nanosystems Initiative Munich (NIM), 80799 M\"unchen, Germany}
\begin{abstract}
Magnetically ordered, electrically insulating materials pave the way towards novel spintronic devices. In these materials the flow of magnetic excitations such as magnons results in pure spin currents. These spin currents can be driven by gradients of the spin chemical potential and/or temperature such that they can play the same role in novel spintronic devices as charge currents in traditional electronic circuits. Connecting spin current based and charge current based devices requires spin to charge interconversion. This has been achieved by the spin Hall effect with an efficiency of several 10\%. The recent progress in materials development and understanding of pure spin current physics opens up a plethora of novel device concepts and opportunities for fundamental studies.
\end{abstract}


\end{frontmatter}


\section{Introduction}

Today's information and communication technology is dominated by electronic and photonic devices. In sensing and magnetic storage applications, however, spin-based devices play an important role. Since many charge-based devices currently approach fundamental performance limits, it is tempting to exploit the spin degree of freedom to surpass these limits. A particular aspect is the reduction of power dissipation. However, since usually the spin (angular momentum) degree of freedom is carried by charge carriers, the flow of spin currents in electrical conductors is inevitably associated with the dissipative flow of charge currents. Aiming at devices with reduced dissipation, an obvious solution is to generate pure spin currents. This can be achieved in magnetically ordered electrical insulators (MOIs). There the charge degree of freedom is frozen out and pure spin currents are obtained by driving a flow spin excitation quanta (magnons) via a gradient of the spin chemical potential or of the temperature~\cite{Tserkovnyak2002,Tserkovnyak2005,Mosendz2010,Ando2010,Uchida2008,Uchida2010,Xiao2010,Bauer2012,Cornelissen2015,Goennenwein2015,nakayama_spin_2013,Vlietstra2013,chen_theory_2013}. The magnons carrying the angular momentum quantum $\hbar$ are the carriers of spin currents in these \emph{spin} conductors, in the same way as electrons carrying the charge quantum $e$ are those of charge currents in \emph{charge} conductors.

In the realm of spintronics, pure spin currents have developed into a new paradigm and are promising for a broad variety of novel device concepts~\cite{Cahaya2015,Chumak2015,Chumak2017}. Over the last decade extensive fundamental research work has been dedicated to find means to efficiently generate and detect pure spin currents as well as to study their transport over length scales relevant for device concepts. In this context it is important to mention that in contrast to charge in an electrical conductor, spin in an angular momentum conductor is not conserved. The angular momentum stored in the magnon system can leak into the phonon system. With an average magnon lifetime $\tau_\mathrm{m}$, we can define a characteristic length $\Lambda_\mathrm{m,diff} = \sqrt{D_\mathrm{m}\tau_\mathrm{m}}$ for diffusive or $\Lambda_\mathrm{m,ball} = v_\mathrm{m}\tau_\mathrm{m}$ for ballistic motion, where $D_\mathrm{m}$ and $v_\mathrm{m}$ are the diffusion constant and group velocity of the magnons, respectively. Evidently, spin current-based devices where magnons are emitted at a source S and collected at a drain D should have a SD-separation much smaller than these characteristic length scales. The latter can be as high as several hundred nanometers in current materials (See Table~\ref{table:magnondiffusion}). Moreover, for transistor-like three-terminal devices suitable implementations of a gate electrode G, allowing for a (dissipationless) control of spin currents are desirable. Here, we briefly review the present understanding of spin current physics, provide an outlook on the future potential of spin current-based devices, and address necessary future research directions with respect to MOIs.

\section{State of the art}

\subsection{Foundations of Pure Spin Current Physics}

Initially, the interest in MOIs was mainly driven by the spin caloritronics community, facing the problem to clearly separate between contributions of charge and spin currents in spin-thermogalvanic phenomena. The obvious solution was to freeze-out the charge contribution by using MOIs. The potential of pure spin currents in MOIs for device applications was rapidly recognized and meanwhile they represent a cornerstone for modern spintronic device concepts. Since our present world is still dominated by electronics, i.e. the charge degree of freedom, there are hardly any commercial sources and detectors available for spin currents. Therefore, the research efforts in spin current physics and devices were focused on the following key tasks:
\begin{enumerate}
\item Generation of pure spin currents, including the efficient conversion of charge currents into spin currents.
\item Detection of pure spin currents, including the efficient conversion of spin into charge currents.
\item Understanding the decay of excess magnons towards thermal equilibrium.
\item Study of the diffusive and ballistic transport of pure spin currents.
\item Development of efficient tools for the control and manipulation of spin currents.
\end{enumerate}

\paragraph*{(1) Generation of pure spin currents:}

Pure spin current generatin schemes include spin pumping (driving non-equilibrium spin dynamics in MOIs by rf-fields~\cite{Tserkovnyak2002,Tserkovnyak2005,Mosendz2010,Ando2010,Heinrich2011,Czeschka2011,Burrowes2012,Hahn2013,Jiao2013} or surface acoustic waves~\cite{Uchida2011,Uchida2012,Weiler2012SPElastic}), driving spin currents by temperature gradients~\cite{Uchida2008,Uchida2010,Xiao2010,Slachter2010,Bauer2012,Weiler2012,Adachi2013,Schreier2013,Rezende2014,Uchida2014} or injecting spin currents into MOIs from a thin normal (not magnetically ordered) metal (NM) deposited on the MOI~\cite{Liu2011,nakayama_spin_2013,althammer_quantitative_2013,Vlietstra2013,Hahn2013}. The latter technique is based on the spin Hall effect (SHE)~\cite{Dyakonov1971,Hirsch1999,Hoffmann2013,Sinova2015} and requires materials with strong spin-orbit coupling (large Berry phase). \add{In this regard, the Berry phase in such materials gives rise to spin-dependent transverse velocities, which in turn give rise to the SHE~\cite{xiao_berry_2010}.} A large variety of MOI/NM heterostructures have been reported and large charge-spin conversion factors (spin Hall angles) exceeding 10\% have been reported in literature for Pt, W, and Ta~\cite{Liu2012Pt,Pai2012,Liu2012}. Moreover, a solid theoretical understanding of the non-equilibrium spin transport across MOI/NM interfaces has been developed~\cite{Bauer2012,Adachi2013,Bender2015}. Unfortunately, the relevant materials parameters such as the spin Hall angle and spin diffusion length in the NM as well as the spin mixing conductance at the MOI/NM interface still show a significant spread in literature, most likely due to considerable variations in materials and interface properties. In particular, the transparency of MOI/NM interfaces for pure spin currents plays a crucial role and requires careful attention in experiments. Fortunately, MOI/NM heterostructures with an interface transparency similar to that in all metallic ferromagnet/NM structures have been achieved, showing that a highly efficient spin current transport across the MOI/NM interface is possible~\cite{Burrowes2012,weiler_experimental_2013}. Early spin pumping experiments mostly focused on the DC part of the spin transport across MOI/NM interfaces. However, more recently also the time-varying part attracted interest as it may pave the way towards high-speed spintronic devices up to the THz~regime~\cite{Kampfrath2013,Hahn2013,Weiler2014,Cheng2014,Cheng2016,Johansen2017}.

In a large number of experiments temperature gradients have been successfully used to drive pure spin currents in MOIs. In open circuits conditions a gradient of the spin chemical potential $\nabla\mu_s$ appears and -- analogous to the charge Seebeck effect -- this phenomenon is named spin Seebeck effect (SSE)~\cite{Uchida2008,Uchida2010,Xiao2010,Slachter2010,Bauer2012,Weiler2012,Adachi2013,Schreier2013,Rezende2014,Uchida2014,Giles2015}. Strictly speaking, the proportionality constant between $\nabla T$ and $\nabla\mu_s$ should be denoted as spin Seebeck coefficient. However, since $\nabla\mu_s$ is difficult to measure, the spin current was converted into a charge current (leading to a voltage in open circuit conditions) using MOI/NM heterostructures (see above) and the Seebeck coefficient has been introduced as the proportionality constant between $\nabla T$ and the detected voltage. An important contribution of SSE experiments was to provide a deeper understanding of magnon excitations in MOIs that were previously only attainable by \add{non-electrical transport} experiments~\cite{Geprgs2016}. Regarding applications, the SSE may allow us to use MOI/NM heterostructures for waste energy reuse in the same way as thermoelectric devices~\cite{Kirihara2012}. The relevant figures of merit are presently under study~\cite{Cahaya2015}. \add{While initial work on the spin Seebeck effect mostly employed ferro- and ferrimagnetic ordered MOIs, within the last years the spin Seebeck effect was also observed in MOIs with antiferromagnetic order (Cr$_2$O$_3$, MnF$_2$)~\cite{seki_thermal_2015,wu_antiferromagnetic_2016,rezende_theory_2016,holanda_spin_2017}. In these experiments large external magnetic fields are required to manipulate the magnetic moments of the antiferromagnet, however they proof that a pure spin current can also be generated by a thermal drive in a MOI with antiferromagnetic order.}

\paragraph*{(2) Detection of pure spin currents:}

Spin currents can be elegantly detected via the inverse spin Hall effect (ISHE), which is the Onsager reciprocal of the SHE. This detection scheme is based on MOI/NM heterostructures where spin currents in the MOI are transferred into a thin NM layer and converted into a charge current by the ISHE. Efficient conversion requires highly spin transparent MOI/NM interfaces and NM layer thicknesses on the lengthscale of the electron spin diffusion length. Typical NM materials are Pt, W, and Ta, featuring spin Hall angles of 10\% or more~\cite{althammer_quantitative_2013,weiler_experimental_2013,Wang2014}. Novel two-dimensional systems approach values of 40\%~\cite{Jamali2015}.

\paragraph*{(3) Decay of excess magnons:}

We already pointed out in the introduction that the relevant magnon transport length-scale crucially depends on the magnon lifetime $\tau_\mathrm{m}$. After switching off an external perturbation, any excess population of a particular magnon with energy $\hbar\omega_m$ and wave number $\mathbf{k}_m$ will decay towards the thermal equilibrium population by magnon-magnon and magnon-phonon interaction. The characteristic decay time $\tau_\mathrm{m}$ will depend on both energy and wave number. The linewidth in ferromagnetic resonance experiments yields access to the uniform ($k=0$) mode of spin excitations. For MOIs long magnon lifetimes (e.g. $\tau_\mathrm{m}=1\;\mathrm{\mu s}$ for yttrium iron garnet [YIG]) exceeding those of metallic ferromagnets by several orders of magnitude have been reported, as the usually dominant magnon-electron scattering is suppressed in MOIs~\cite{Sparks1964,Gurevich1996}. The decay of finite wave number magnons can be studied by optical techniques which, however, are \add{mostly} restricted to small wave numbers~\cite{Demokritov2001,Demidov2007,Demidov2008,Demokritov2008,Sebastian2015}. \add{These inelastic light scattering experiments rely on the inductive excitation of magnons via microwave antenna and spatially resolved experiments allow to investigate magnon transport in MOIs.} A detailed understanding of the magnon relaxation as a function of energy and wave number is still missing and requires further research efforts in both theory and experiment.

\paragraph*{(4) Diffusive and ballistic transport of pure spin currents:}

\begin{figure}[tbh]
\centering
 \includegraphics[width=0.8\columnwidth]{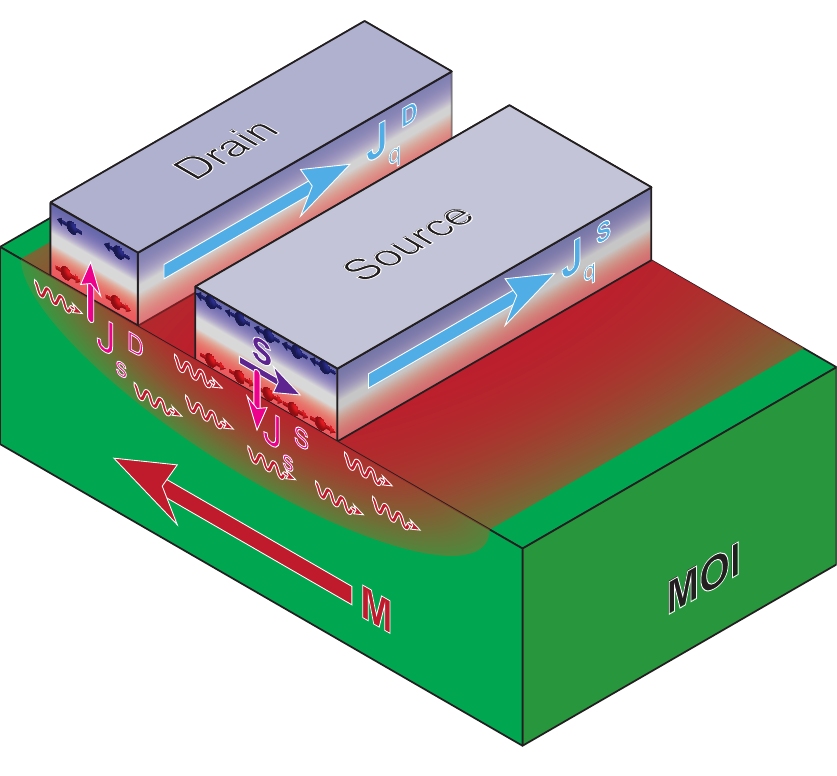}
 \caption[Experimental configuration for the study of spin transport in MOIs]{Illustration of the experimental configuration for the study of spin transport in MOIs. Two NM strips are placed on a magnetically ordered insulator (MOI) at distance $d$ and act as the source S and drain/detector D for a magnon (spin) current, diffusively or ballistically propagating between S and D. Further explanations are given in the text. }
  \label{Figure_MMR}
\end{figure}

The realization of both charge-to-spin and spin-to-charge current conversion not only allowed for the study of long-distance magnon transport in MOI/NM heterostructures, but also laid the foundations for interfacing charge and spin based devices and realizing magnon based logic~\cite{Cornelissen2015,Cornelissen2016,Cornelissen2016_fielddependence,Cornelissen2016_temperaturedependence,Shan2017_NFO,Liu2017,Goennenwein2015,Ganzhorn2016,Ganzhorn2017,Vlez2016,Li2016_Riverside}. To study magnon transport in MOIs, two NM strips -- one acting as the magnon source (S) and the other as the magnon drain (D) -- are placed with separation $d$ on one and the same MOI as shown in Fig.~\ref{Figure_MMR}~\cite{Zhang2012,Zhang2012_PRB,Cheng2017}. The magnons generated via the SHE in S diffusively or ballistically propagate to D, where they are converted into a voltage signal via the ISHE. The size of the voltage signal crucially depends on temperature, applied magnetic field, the SD-separation, and the magnetization direction in the MOI relative to the S and D strips, leading to the magnon-mediated magnetoresistance~\cite{Cornelissen2015,Goennenwein2015}.

In the following we briefly summarize the physical processes relevant for the experimental configuration shown in Fig.~\ref{Figure_MMR}. First, a charge current $\mathbf{J}_\mathrm{q}^\mathrm{S}$ flowing along S is converted into a spin current $\mathbf{J}_\mathrm{s}^\mathrm{S}$ by the spin Hall effect. $\mathbf{J}_\mathrm{s}^\mathrm{S}$ has a polarization $\mathbf{s}$ and is flowing \add{across the interface}. Second, this spin current generates a steady state electron spin accumulation in the NM at the MOI/NM interface. Third, by electron-magnon scattering at the MOI/NM interface the spin accumulation is transferred into the MOI, corresponding to the injection of a spin current into the MOI. For the further discussion we have to distinguish between the two cases $\mathbf{M} \perp \mathbf{s}$ and $\mathbf{M} \| \mathbf{s}$. For $\mathbf{M} \perp \mathbf{s}$, the spin current injected into the MOI is rapidly dissipated by spin transfer torque and no long-distance magnon transport is observable. That is, we expect a drain/detector voltage $V_d=0$ for $\mathbf{M} \perp \mathbf{s}$. In contrast, for $\mathbf{M} \| \mathbf{s}$ spin transfer torque is negligible. In this case a magnon accumulation or depletion is generated at the MOI/NM interface for $\mathbf{M}$ antiparallel and parallel to $\mathbf{s}$, respectively. In the fourth step, this accumulation/depletion will diffusively or ballistically propagate to the drain/detector strip, with details of this transport process depending on temperature, magnon density and scattering rates. Fifth, the magnon propagation will generate a magnon accumulation/depletion at the MOI/NM interface of drain strip D. Finally, inversely to the source strip S, by electron-magnon scattering this accumulation/depletion will be transferred into the electron spin system at the MOI/NM interface. This leads to a spin current $\mathbf{J}_\mathrm{s}^\mathrm{D}$ in NM which is then converted into a charge current $\mathbf{J}_\mathrm{q}^\mathrm{D}$ via the ISHE and measured as a voltage in open circuit conditions. Since an electrical current in S causes a voltage drop in the spatially separated D, the phenomenon has been called nonlocal magnetoresistance or, as it is mediated by magnons, magnon mediated magnetoresistance (MMR)~\cite{Cornelissen2015,Goennenwein2015}.

The experimental configuration shown in Fig.~\ref{Figure_MMR} has been successfully used to study diffusive magnon transport in YIG thin films grown by liquid phase epitaxy (LPE) and to determine the magnon diffusion length $\Lambda_\mathrm{m,diff}$. In the first experimental work by Cornelissen~\textit{et al.}~\cite{Cornelissen2015} an astonishingly large $\Lambda_\mathrm{m,diff}=9,400\;\mathrm{nm}$ was found in LPE-YIG at room temperature. This value is very promising for device applications, since the maximum SD separation is limited by $\Lambda_\mathrm{m,diff}$. However, values varying by more than two orders of magnitude have been found in subsequent experiments including sputtered YIG and nickel ferrite films (cf. Table~\ref{table:magnondiffusion}), indicating that the measured numbers strongly depend on materials properties and details of the device design including the thickness of the MOI~\cite{Liu2017}, fabrication process and data analysis. When comparing and interpreting the measured magnon diffusion lengths, one has to keep in mind that several magnon branches and magnons with different wave number are contributing~\cite{Cherepanov1993}. This is related to the fact that the magnons are generated by electron-magnon scattering at the MOI/NM interface, resulting in a broad excitation spectrum~\cite{Zhang2012,Zhang2012_PRB,Bender2015,Cornelissen2016_temperaturedependence}. Hence, the measured $\Lambda_\mathrm{m,diff}$ represents an average value obtained for specific experimental conditions. \add{In contrast, experiments using inductive excitation mechanisms can selectively excite certain magnon wavelengths yielding rather different propagation lengths as to the non-local experiments (See Table~\ref{table:magnondiffusion}). Another indirect measurement method may be provided by the thermal effects associated to the magnetic excitation transport~\cite{onose_observation_2010,an_unidirectional_2013}, but requires further experimental studies.} Reducing the temperature to $T\leq30\;\mathrm{K}$ was found to result in a vanishingly small detector voltage. This has been interpreted as evidence for a thermally activated nature of the magnon generation process at the MOI/NM interface~\cite{Cornelissen2016_temperaturedependence}. Meanwhile, the initial experiments have been confirmed by other groups and led to a deeper understanding of the underlying physics~\cite{Cornelissen2015,Cornelissen2016,Cornelissen2016_fielddependence,Cornelissen2016_temperaturedependence,Shan2017_NFO,Liu2017,Goennenwein2015,Ganzhorn2016,Ganzhorn2017,Vlez2016,Li2016_Riverside}. \add{Interestingly, amorphous YIG thin films also exhibit long-distance magnon-transport in non-local experiments, which may allow to even use disordered spin systems as materials for magnon devices~\cite{wesenberg_long-distance_2017}.}

\begin{table}
\begin{tabular}{|l|c|c|c|c|}
  \hline
  material & $\Lambda_\mathrm{m,diff} \;\mathrm{(nm)}$ & $t_{\mathrm{MOI}} \;\mathrm{(nm)}$ & method & Ref. \\
  \hline
  LPE-YIG & $31,000$ & $100$ & optical& \cite{pirro_spin-wave_2014} \\
  YIG, sput. & $10,060$ & $40$ & optical& \cite{jungfleisch_spin_2015} \\
  YIG, PLD & $25,000$ & $20$ & optical& \cite{collet_spin-wave_2017} \\
  LPE-YIG & $860,000$ & $200$ & inductive& \cite{maendl_spin_2017} \\
  LPE-YIG & $9,400$ & $100$ & lateral & \cite{Cornelissen2015} \\
  LPE-YIG & $700$ & $3,000$ & lateral & \cite{Goennenwein2015} \\
  YIG, sput. & 38 & $40$ to $100$ & vertical & \cite{Wu2016_SputteredYIGVertical} \\
  NiFe$_2$O$_4$, sput. & $3,100$ & $44$ & lateral & \cite{Shan2017_NFO} \\
  \hline
\end{tabular}
\caption[Magnon diffusion length measured in various experiments]{Magnon diffusion lengths $\Lambda_\mathrm{m,diff}$ measured in different experiments for different MOIs with different thicknesses $t_{\mathrm{MOI}}$ of the MOI at $T=300\;\mathrm{K}$. \add{The different methods used are: spatially resolved inelastic light scattering (optical), inductive magnon generation and detection (inductive), lateral (lateral) and vertical (vertical) non-local electrical transport measurements.}}
\label{table:magnondiffusion}
\end{table}

A further crucial parameter is the so-called spin convertance describing the efficiency of the spin transfer between the electron and magnon system at the MOI/NM interface~\cite{Zhang2012,Zhang2012_PRB}. Very recently, means to enhance the spin convertance have been proposed~\cite{Vlez2016}. In summary, one can say that the measured magnon diffusion lengths in MOIs are promising, but more systematic experiments are required to gain a full quantitative understanding.

\paragraph*{(5) Tools for the control and manipulation of spin currents:}

Three-terminal, transistor-like devices are highly important for applications. Therefore, regarding the implementation of device concepts not only the generation and detection of spin currents but also the control and manipulation of spin current flow by suitable gate electrodes is required. Research in this direction is only rudimentary and requires more attention. Current approaches are mostly centered on employing reconfigurable magnonic crystals~\cite{Chumak2015,Chumak2017}. \add{Another pathway towards spin current control in MOIs is to employ nonlinear magnon-magnon interactions, which can be used for example to suppress a single wavevector magnon current~\cite{chumak_magnon_2014}. However, this approach requires large local magnon densities, which may be achieved by making use of the magnon dispersion in magnonic crystals. Ideally, similar to charge based transistor concepts an electrical field control magnon conduction channel would allow for very efficient magnon logic devices.}

\subsection{Novel Magnetoresistive Effects}

The interplay between the electron spin accumulation generated in a NM with strong spin-orbit interaction via the SHE and the non-equilibrium magnon distribution in a MOI at an MOI/NM interface led to novel magnetoresitance phenomena, namely the spin Hall magnetoresistance (SMR) and the magnon mediated magnetoresistance (MMR), which we discuss in more detail in the following.

\paragraph*{Spin Hall Magnetoresistance (SMR):}

The combination of SHE and ISHE for pure spin current generation and detection in MOI /NM heterostructures led to the discovery of the spin Hall magnetoresistance (SMR)~\cite{nakayama_spin_2013,Vlietstra2013,Hahn2013SMR,chen_theory_2013}. The SMR is characterized by the dependence of the NM resistance on the angle between charge current direction in the NM and the (sublattice) magnetization orientation in the MOI. In the simplest case of a collinear ferromagnet, an additional contribution to the resistance in the NM arises if the spin polarization $\mathbf{s}$ of the spin current generated in the NM via the SHE and transferred across the MOI/NM interface is perpendicular to the magnetization $\mathbf{M}$ in the MOI. For $\mathbf{s} \| \mathbf{M}$, the dissipation channel resulting from spin transfer torque and, hence, the additional NM resistance is absent. The magnitude of the SMR effect scales with the square of the spin Hall angle of the NM layer, i.e. the interconversion factor between charge and spin current. Experimental values for the relative resistance change of up to $10^{-3}$ have been reported~\cite{althammer_quantitative_2013,weiler_experimental_2013}. Most importantly, from these values, which can be obtained from simple magnetotransport measurements, relevant spin transport parameters can be inferred. Furthermore, SMR proved to be a powerful tool also for the investigation of non-collinear magnetic phases (e.g.~helical, spin-canting and spin-flip ordering) in MOIs~\cite{Aqeel2015,Ganzhorn2016SMR}. \add{Especially MOIs with antiferromagnetic order have been investigated using the SMR~\cite{han_antiferromagnet-controlled_2014,manchon_spin_2017,hou_tunable_2017,ji_spin_2017,hoogeboom_negative_2017,fischer_spin_2017}, which further provides evidence that the SMR is sensitive to the magnetic order parameter itself (or possibly the magnetic sublattice moment orientations), but not the net magnetic moment direction.} Taken things even a step further, spin currents in NM have meanwhile been used to switch the magnetization of a MOI by the spin transfer torque~\cite{Avci2016}.

\paragraph*{Magnon Mediated Magnetoresistance (MMR):}

As discussed above, in a device structure as shown in Fig.~\ref{Figure_MMR} an electrical current in the source S causes a non-local voltage drop in the spatially separated drain D. Since this phenomenon is based on magnon diffusion between S and D, it is called magnon mediated magnetoresistance (MMR). While the theoretical prediction of MMR was pioneered by Zhang and Zhang~\cite{Zhang2012,Zhang2012_PRB} in 2012, the first experimental observation was reported in 2015~\textit{et al.}~\cite{Cornelissen2015,Goennenwein2015}. Although a detailed knowledge of MMR has not yet been achieved, it is evident already today that MMR has a broad potential for magnon based spintronic devices. Due to a long magnon diffusion length in MOIs lateral device schemes as illustrated in Fig.~\ref{Figure_MMR} are feasible with low demands on lateral dimensions. Making use of MMR, electrically controlled magnon logic circuits~\cite{Ganzhorn2016} have already been demonstrated and, very recently, evidence for the magnon transport analogue of the anisotropic magneto resistance in MOIs has been reported~\cite{Liu2017}.

\section{Perspectives for Magnetic Insulator based Spintronics}

Even today, a rich variety of physical phenomena has been studied in MOIs and successfully used for spintronics applications. Among them are the SMR and MMR, which are powerful tools for probing spin transport across MOI/NM interfaces and magnon propagation in MOIs. MMR already has been successfully  used to implement charge-to-spin current converters and magnon-based logic devices~\cite{Ganzhorn2016}. In the following we address some important tasks and key challenges for future research in MOI based spintronics.

\paragraph*{Magnon-Based Transistors:}

An important task will be the realization of transistor-like structures relying on magnon flow in a MOI controlled by a gate electrode. Here, the key challenge is to develop suitable gate concepts. Controlling magnon flow by an electric gate voltage would be particularly desirable in the context of electronic-magnonic hybrid circuits. Here, multiferroic materials with long magnon diffusion lengths are an option. Research in this direction is expected to also stimulate the realization of novel magnon logic devices.

\paragraph*{Selective Magnon Excitation:}

Presently, there is little control on magnon excitation in devices relying on the SHE in MOI/NM heterostructures. The development of concepts allowing to selectively excite magnons of specific energy and wave number are a prerequisite for coherent magnon logic circuits using a DC excitation schemes. Possible solutions for selective magnon excitation could be based on magnonic crystals, i.e. on appropriately engineering the magnonic band structure of the MOI~\cite{Chumak2017}.

\paragraph*{Nonlinear Phenomena:}

Another important research task will be the study of nonlinear effects in MOI based devices. A particularly interesting example is the nonlinear dependence of the magnon population in an MOI on the charge current flowing in a NM in devices based on MOI/NM heterostructures. First experiments show that the excitation of magnons with low damping is possible after overcoming a threshold value~\cite{thiery_nonlinear2017}. Similarly, the generation of magnonic superfluids using spin currents generated via the SHE are already has been proposed~\cite{Takei2015_SpinSuperfluid}. While Bose-Einstein condensation in MOIs has already been experimentally realized using parametric microwave pumping~\cite{Demokritov2006,Demidov2007}, an experimental realization of a magnonic BEC by all-electrical driving has not yet been reported. We note that a magnonic superfluid would coexist with normal magnonic excitations, leading to short-circuiting effects of the magnon transport in MOIs. This allows for novel control schemes for magnonic transport.

\paragraph*{Antiferromagnetic Magnonics:}

Current experimental studies mostly focus on magnon transport in ferro- or ferrimagnetically ordered MOIs, where typical magnon frequencies are of the order of a few GHz. However, these studies and already developed device concepts can be straightforwardly transferred to antiferromagnetically ordered MOIs \add{as already demonstrated for the spin Seebeck effect~\cite{seki_thermal_2015,wu_antiferromagnetic_2016,rezende_theory_2016,holanda_spin_2017} and for the spin Hall magnetoresistance~\cite{han_antiferromagnet-controlled_2014,manchon_spin_2017,hou_tunable_2017,ji_spin_2017,hoogeboom_negative_2017,fischer_spin_2017}}, thereby extending magnon frequencies to the THz regime~\cite{Cheng2014,Cheng2016,Johansen2017}. It is expected that this leads to promising applications with an extended frequency regime. \add{A promising example for future applications is the recent demonstration of an antiferromagnetic memory based on the manipulation by the magnetoelectric effect in Cr$_2$O$_3$ thin films and readout via the anomalous Hall effect in a Pt electrode~\cite{kosub_purely_2017}. Clearly, further studies to better understand retainability, stability, repeatability and switching speed are required for successful implementation into new device concepts.}

\paragraph*{Ballistic Magnon Transport:}

Until today, magnon based devices mostly rely on diffusive magnon transport. An important future task will be the detailed study of ballistic magnon transport. A particulary interesting aspect is the study of ballistic magnon transport in one-dimensional transport channels and the demonstration of the quantized magnon conductance in such structures. The experimental challenge is to fabricate one-dimensional magnon transport channels and to reduce the length of these channels below the magnon mean free path. To this end, a possible strategy is to use domain walls in MOIs. We note that the realization of ballistic magnon transport also is promising regarding novel device concepts.

Taken together, within the last decade the successful research on spin current transport in MOIs has lead to a manifold of novel phenomena. This allowed us to get a deeper insight into the fundamental physics and paved the way towards promising applications. Over the next years, further exciting results on fundamental physics aspects and rapid progress regarding transfer into applications are expected.




\bibliographystyle{elsarticle-num}
\bibliography{Biblio}
%
%
%
\end{document}